\documentclass[proceedings]{JHEP3}
\usepackage{amsfonts}
\usepackage{amsmath}
\usepackage{epsfig}

\newbox\mybox

\newcommand\fverb{\setbox\mybox=\hbox\bgroup\verb}
\newcommand\fverbdo{\egroup\medskip\noindent\fbox{\unhbox\mybox}\ }
\newcommand\fverbit{\egroup\item[\fbox{\unhbox\mybox}]}
\conference{Squeezed coherent states for noncommutative spaces}
\abstract{We provide an explicit construction for Gazeau-Klauder coherent states related to non-Hermitian Hamiltonians
with discrete bounded below and nondegenerate eigenspectrum. The underlying spacetime structure is taken to be of a 
noncommutative type with associated uncertainty relations implying minimal lengths. The uncertainty relations for the constructed states 
are shown to be saturated in a Hermitian as well as a non-Hermitian setting for a perturbed harmonic oscillator. The computed 
value of the Mandel parameter dictates that the coherent wavepackets are assembled according to sub-Poissonian statistics.
Fractional revival times, indicating the superposition of classical-like sub-wave packets are clearly identified.}

\title{Squeezed coherent states for noncommutative spaces with minimal
length uncertainty relations}
\author{Sanjib Dey and Andreas Fring \\
Centre for Mathematical Science, City University London,\\
Northampton Square, London EC1V 0HB, UK\\
E-mail: sanjib.dey.1@city.ac.uk, a.fring@city.ac.uk}

\input{tcilatex}
\begin{document}

\section{Introduction}

Noncommutative spacetime structures are suggested by gravitational stability 
\cite{dop} in almost all promising approaches to quantum gravity, such as
string theory \cite{String1,String2,wit} or loop quantum gravity \cite%
{Ashtekar,Rovelli} as well as black hole physics \cite{Scardigli}. Besides
the numerous possible structures we focus here on a particularly interesting
one giving rise to minimal measurable distance beyond which the entire
concept of length becomes meaningless. Such type of cutoff in our possible
knowledge of space results from generalized versions of Heisenberg's
uncertainty relations, i.e. from modifying standard spacetime to certain
noncommutative versions in a specific way. These type of spaces have
attracted considerable attention in recent years in the mathematical and
physical literature \cite%
{Kempf1,Kempf2,AFBB,AFLGFGS,AFLGBB,DFG,Aschieri:2005zs,Gomes:2009tk,Dorsch:2011qf,Castro:2011in,Chang:2011jj,Pedram:2011xj,Nozari:2012gd,Maz}%
.

So far most of the attention has been paid to the algebras, but fairly
little, in comparison to standard systems, to the actual nature and
properties of the states it acts on. Here we focus on the explicit
construction of coherent states respecting these modified uncertainty
relations. Generic expressions for coherent states were propose by Gazeau
and Klauder (GK) \cite{Klauder1,Klauder2}. By construction the states are
expected to be stable when evolved in time, in the sense that they remain
coherent during the evolution process. For a one dimensional harmonic
oscillator on a noncommutative space we present here a computation of these
states in first order perturbation theory. When dealing with noncommutative
spacetime structures of the aforementioned type one also encounters an
addition complication due to the fact that the associated canonical
variables are in general no longer Hermitian with regard to standard inner
products \cite{AFBB}. Consequently Hamiltonians formulated in terms of these
variables are also no longer Hermitian. We adopt here a recent approach to
non-Hermitian systems \cite{Urubu,Bender:1998ke,AliI,Benderrev,Alirev} which
renders them self-consistent and physically meaningful.

A striking feature of the coherent states presented here is the well known
fact that in a certain parameter regime the original wave packet can be
fully reconstructed after a specific time and can be interpreted as a
superposition of classical-like sub-wave packets. In the system considered
here the existence of these structures is a signature of the spacetime
deformation and disappears in a standard setting. This indicates the
interesting possibility that the structures of noncommutative spaces could
actually be probed experimentally.

Many of the computations presented here have been attempted previously by
Ghosh and Roy in \cite{GhoshRoy}, but almost all our results disagree with
their findings and conclusions which we believe to be conceptually and
computationally incorrect as we will point out in the course of our
presentation.

Our manuscript is organized as follows: In section 2 we assemble various
generalities on GK-coherent states and show how the construction needs to be
altered for a non-Hermitian setting. We also set up our notation for a
standard perturbative treatment. In section 3 we construct GK-states and
evaluates various expectation values for a Hermitian as well as
non-Hermitian setting, which we use to test the Ehrenfest theorem and the
equivalence principle. In section 4 we present our analysis for the revival
times structure. Our conclusions are stated in section 5.

\section{Perturbative GK-coherent states for non-Hermitian Hamiltonians}

We commence by establishing our notations by collecting some well known
facts about GK-coherent states and indicate the necessary modifications
needed for a non-Hermitian setting. The GK-coherent states \cite%
{Klauder1,Klauder2} for a Hermitian Hamiltonian $h$ with discrete bounded
below and nondegenerate eigenspectrum are defined as a two parameter set 
\begin{equation}
\left\vert J,\gamma ,\phi \right\rangle =\frac{1}{\mathcal{N}(J)}%
\sum\limits_{n=0}^{\infty }\frac{J^{n/2}\exp (-i\gamma e_{n})}{\sqrt{\rho
_{n}}}\left\vert \phi _{n}\right\rangle ,\text{\qquad }J\in \mathbb{R}%
_{0}^{+},\gamma \in \mathbb{R}.  \label{GK}
\end{equation}%
The states $\left\vert \phi _{n}\right\rangle $ are the orthonormal
eigenstates of $h$, that is $h\left\vert \phi _{n}\right\rangle =\hbar
\omega e_{n}\left\vert \phi _{n}\right\rangle $. The probability
distribution and normalization constant are%
\begin{equation}
\rho _{n}:=\dprod\limits_{k=1}^{n}e_{k}\qquad \text{and\qquad }\mathcal{N}%
^{2}(J):=\dsum\limits_{k=0}^{\infty }\frac{J^{k}}{\rho _{k}},  \label{N2}
\end{equation}%
respectively, where the latter results from the requirement $\left\langle
J,\gamma ,\phi \right. \!\!$ $\left\vert J,\gamma ,\phi \right\rangle =1$.

Here we will consider Hamiltonians $H$ on noncommutative spaces which are
non-Hermitian with regard to the standard inner product. The construction
for the coherent states is then easily adoptable when we assume that the
Hamiltonian $H$ is pseudo/quasi Hermitian, i.e. the non-Hermitian
Hamiltonian $H$ and the Hermitian Hamiltonian $h$ are related by a
similarity transformation $h=\eta H\eta ^{-1}$, with $\eta ^{\dagger }\eta $
being a positive definite operator playing the role of the metric. The
corresponding eigenstates $\left\vert \Phi \right\rangle $ and $\left\vert
\phi \right\rangle $ of $H$ and $h$, respectively, are then simply related
as $\left\vert \Phi \right\rangle $ $=\eta ^{-1}\left\vert \phi
\right\rangle $. As observables are expected to be Hermitian, we need to
change the metric when computing expectation values for operators associated
to the non-Hermitian system \cite{Urubu,Bender:1998ke,AliI,Benderrev,Alirev}%
. The same reasoning has to be adopted for the evaluation of expectation
values with regard to the coherent states. Therefore the expectation value
for a non-Hermitian operator $\mathcal{O}$ related to a Hermitian operator $%
o $ by a similarity transformation $o=\eta \mathcal{O}\eta ^{-1}$ is
computed as%
\begin{equation}
\left\langle J,\gamma ,\Phi \right\vert \mathcal{O}\left\vert J,\gamma ,\Phi
\right\rangle _{\eta }:=\left\langle J,\gamma ,\Phi \right\vert \eta
^{\dagger }\eta \mathcal{O}\left\vert J,\gamma ,\Phi \right\rangle
=\left\langle J,\gamma ,\phi \right\vert o\left\vert J,\gamma ,\phi
\right\rangle .  \label{O}
\end{equation}%
Our notation is to be understood in the sense that in the state $\left\vert
J,\gamma ,\phi \right\rangle $ and $\left\vert J,\gamma ,\Phi \right\rangle $
we sum over the eigenstates of the Hermitian Hamiltonian $h$ and
non-Hermitian Hamiltonian $H$, respectively. These states are continuous in
the two variables $(J,\gamma )$, provide a resolution of the identity, are
temporarily stable, in the sense that they remain coherent states under time
evolution, and satisfy the action identity%
\begin{equation}
\left\langle J,\gamma ,\Phi \right\vert H\left\vert J,\gamma ,\Phi
\right\rangle _{\eta }=\left\langle J,\gamma ,\phi \right\vert h\left\vert
J,\gamma ,\phi \right\rangle =\hbar \omega J.  \label{AI}
\end{equation}%
This identity ensures that $(J,\gamma )$ are action angle variables \cite%
{Klauder1,Klauder2}.

The main purpose is here to consider a model on a noncommutative space with
nontrivial commutation relations for their canonical variables giving rise
to minimal uncertainties. It is then interesting to investigate how close
the GK-states approach the minimum uncertainty product and eventually might
even become squeezed states. Thus for a simultaneous measurement of two
observables $A$ and $B$ in this system we need to evaluate the left and
right hand side of the generalized version of Heisenberg's uncertainty
relation 
\begin{equation}
\Delta A\Delta B\geq \frac{1}{2}\left\vert \left\langle J,\gamma ,\Phi
\right\vert [A,B]\left\vert J,\gamma ,\Phi \right\rangle _{\eta }\right\vert
.  \label{HU}
\end{equation}%
The uncertainties are computed as $\Delta A=\left\langle J,\gamma ,\Phi
\right\vert A^{2}\left\vert J,\gamma ,\Phi \right\rangle _{\eta
}-\left\langle J,\gamma ,\Phi \right\vert A\left\vert J,\gamma ,\Phi
\right\rangle _{\eta }^{2}$ and analogously for $\Delta B$. In order to test
the quality of the coherent states, i.e. to see how closely they resemble
classical mechanics, we may also test Ehrenfest's theorem for an operator $A$
\begin{equation}
i\hbar \frac{d}{dt}\left\langle J,\gamma +t\omega ,\Phi \right\vert
A\left\vert J,\gamma +t\omega ,\Phi \right\rangle _{\eta }=\left\langle
J,\gamma +t\omega ,\Phi \right\vert [A,H]\left\vert J,\gamma +t\omega ,\Phi
\right\rangle _{\eta }.  \label{Ehren}
\end{equation}%
We used in (\ref{Ehren}) the fact that the time evolution for the states $%
\left\vert J,\gamma ,\Phi \right\rangle $ is simply implemented as $\exp
(-iHt/\hbar )\left\vert J,\gamma ,\Phi \right\rangle =\left\vert J,\gamma
+t\omega ,\Phi \right\rangle $, see \cite{Klauder1,Klauder2}. Specifying the
operators $A$ and $B$ we will also test below the correspondence principle.

Here we present a perturbative treatment of the above considerations around $%
h_{0}$ for a Hamiltonian decomposable as $h=h_{0}+h_{1}$, with $%
h_{0}\left\vert n\right\rangle =e_{n}^{(0)}\left\vert n\right\rangle $.
According to standard Rayleigh-Schr\"{o}dinger perturbation theory the first
order expansions of the eigenenergies and the eigenstates are%
\begin{equation}
e_{n}=e_{n}^{(0)}+\left\langle n\right\vert h_{1}\left\vert n\right\rangle +%
\mathcal{O}(\tau ^{2})\quad \text{and\quad }\left\vert \phi
_{n}\right\rangle =\left\vert n\right\rangle +\dsum\limits_{k\neq n}\frac{%
\left\langle k\right\vert h_{1}\left\vert n\right\rangle }{%
e_{n}^{(0)}-e_{k}^{(0)}}\left\vert k\right\rangle +\mathcal{O}(\tau ^{2}),
\label{ersteO}
\end{equation}%
respectively. Wherever appropriate, we then simply use these expressions in (%
\ref{GK}) for our computations.

\section{GK-coherent states for the noncommutative harmonic oscillator}

We will now construct the GK-coherent states and various expectation values
for the one dimensional harmonic oscillator \cite{Kempf1,AFBB,DFG} 
\begin{equation}
H=\frac{P^{2}}{2m}+\frac{m\omega ^{2}}{2}X^{2}-\hbar \omega \left( \frac{1}{2%
}+\frac{\tau }{4}\right)   \label{HD1}
\end{equation}%
defined on the noncommutative space satisfying 
\begin{equation}
\lbrack X,P]=i\hbar \left( 1+\check{\tau}P^{2}\right) ,\qquad X=(1+\check{%
\tau}p^{2})x,\qquad P=p.  \label{one}
\end{equation}%
Here $\check{\tau}:=\tau /(m\omega \hbar )$ has the dimension of an inverse
squared momentum with $\tau ~$being dimensionless. We also provided in (\ref%
{one}) a representation for the noncommutative variables in terms of the
standard canonical variables $x$, $p$ satisfying $[x,p]=i\hbar $. The ground
state energy is conveniently shifted to allow for a factorization of the
energy. The Hamiltonian in (\ref{HD1}) in terms of $x$, $p$ differs from the
one treated recently in \cite{GhoshRoy} as we take a different
representation for $X$ and $P$, which we believe to be incorrect in \cite%
{GhoshRoy} even up to $\mathcal{O}(\tau )$. The so-called Dyson map $\eta $,
whose adjoint action relates the non-Hermitian Hamiltonian in (\ref{HD1}) to
its isospectral Hermitian counterpart $h$, is easily found to be $\eta =(1+%
\check{\tau}p^{2})^{-1/2}$. With the help of this expression we evaluate%
\begin{equation}
h=\eta H\eta ^{-1}=\frac{p^{2}}{2m}+\frac{m\omega ^{2}}{2}x^{2}+\frac{\omega
\tau }{4\hbar }(p^{2}x^{2}+x^{2}p^{2}+2xp^{2}x)-\hbar \omega \left( \frac{1}{%
2}+\frac{\tau }{4}\right) +\mathcal{O}(\tau ^{2}).  \label{h}
\end{equation}%
Taking now $h_{0}$ to be the standard harmonic oscillator, the energy
eigenvalues for $H$ and $h$ were computed to lowest order in perturbation
theory \cite{Kempf1,DFG} to 
\begin{equation}
E_{n}=\hbar \omega e_{n}=\hbar \omega n\left[ 1+\frac{\tau }{2}(1+n)\right] +%
\mathcal{O}(\tau ^{2}).  \label{En}
\end{equation}%
According to (\ref{ersteO}) we now calculate the first order expression for
the wavefunctions to%
\begin{equation}
\left\vert \phi _{n}\right\rangle =\left\vert n\right\rangle -\frac{\tau }{16%
}\sqrt{(n-3)_{4}}\left\vert n-4\right\rangle +\frac{\tau }{16}\sqrt{(n+1)_{4}%
}\left\vert n+4\right\rangle +\mathcal{O}(\tau ^{2}).  \label{phi}
\end{equation}%
where $(x)_{n}:=\Gamma (x+n)/\Gamma (x)$ denotes the Pochhammer symbol. We
stress that it is vital to include the second and third term in $\left\vert
\phi _{n}\right\rangle $ in order to achieve an accuracy of order $\tau $
for expectation values. In \cite{GhoshRoy}, where a similar computation was
attempted, these terms were incorrectly ignored. The expression for $E_{n}$
coincides with the one found in \cite{GhoshRoy} for $\tau \rightarrow
2\lambda $, as this computation only involves $\left\langle k\right\vert
h_{1}\left\vert n\right\rangle $. Given $e_{n}$ as defined by by the
relation (\ref{En}), we compute the probability density and the expansions
of its inverse%
\begin{equation}
\rho _{n}=\frac{1}{2^{n}}\tau ^{n}n!\left( 2+\frac{2}{\tau }\right)
_{n}\qquad \text{and\qquad }\frac{1}{\rho _{n}}=\frac{1}{n!}-\frac{3+n}{%
4(n-1)!}\tau +\mathcal{O}(\tau ^{2}),
\end{equation}%
We use the latter expression to evaluate the normalization constant in (\ref%
{N2}) 
\begin{equation}
\mathcal{N}^{2}(J)=e^{J}\left( 1-\tau J-\frac{\tau }{4}J^{2}\right) +%
\mathcal{O}(\tau ^{2}).
\end{equation}%
We have now assembled all the necessary quantities to define the GK-coherent
states $\left\vert J,\gamma ,\phi \right\rangle $ in (\ref{GK}) and are in
the position to verify the validity of some of the crucial requirements on
them, test their behaviour and compute expectation values.

As is well known \cite{Urubu,Bender:1998ke,AliI,Benderrev,Alirev}, in a
non-Hermitian setting the observables are not dictated by the Hamiltonian
and it becomes a matter of choice to select them or equivalently the metric 
\cite{Urubu}. In fact, this is also true for a Hermitian system, where,
however, the choice of the standard metric seems to be the most natural one.
Here we are mainly interested in the Hamiltonian $H$ of (\ref{HD1}) with $X$
and $P$ as observables, but it will also be instructive to consider first
the Hermitian system described by $h$ with $x$ and $p$ being the observables
of choice.

\subsection{Obervables in the Hermitian system}

At first we consider the Hamiltonian $h$ in (\ref{h}) as fundamental and
treat the variables $x$ and $p$ as observables in that system. Expectation
values are then most easily computed by taking the states $\left\vert
n\right\rangle $ to be the normalized standard Fock space eigenstates of the
harmonic oscillator with usual properties $a^{\dagger }\left\vert
n\right\rangle =\sqrt{n+1}\left\vert n+1\right\rangle $ and $a\left\vert
n\right\rangle =\sqrt{n}\left\vert n-1\right\rangle $. To first order in $%
\tau $, we then compute the expectation values of the creation and
annihilation operators%
\begin{eqnarray}
\left\langle J,\gamma ,\phi \right\vert a\left\vert J,\gamma ,\phi
\right\rangle  &=&\sqrt{J}e^{-i\gamma }\left[ 1-\frac{\tau }{4}\left(
2+J+4i\gamma (1+J)\right) \right] +\frac{\tau }{4}J^{3/2}e^{3i\gamma }+%
\mathcal{O}(\tau ^{2}),~~  \label{a} \\
\left\langle J,\gamma ,\phi \right\vert a^{\dagger }\left\vert J,\gamma
,\phi \right\rangle  &=&\sqrt{J}e^{i\gamma }\left[ 1-\frac{\tau }{4}\left(
2+J-4i\gamma (1+J)\right) \right] +\frac{\tau }{4}J^{3/2}e^{-3i\gamma }+%
\mathcal{O}(\tau ^{2}).  \label{ad}
\end{eqnarray}%
For the details of the computation we refer the reader to the appendix. In
what follows we will often drop the explicit mentioning of the order in $%
\tau $, understanding that all our computations are carried out to first
order. Using the fact that $x=\sqrt{\hbar /(2m\omega )}(a^{\dagger }+a)$ and 
$p=i\sqrt{\hbar m\omega /2}(a^{\dagger }-a)$, the expectation values%
\begin{eqnarray}
\left\langle J,\gamma ,\phi \right\vert x\left\vert J,\gamma ,\phi
\right\rangle  &=&\sqrt{\frac{2J\hbar }{m\omega }}\left[ \cos \gamma -\tau %
\left[ \gamma \sin \gamma +\frac{\cos \gamma }{2}+J\sin \gamma \left( \gamma
+\frac{\sin 2\gamma }{2}\right) \right] \right] ,  \label{xp} \\
\left\langle J,\gamma ,\phi \right\vert p\left\vert J,\gamma ,\phi
\right\rangle  &=&-\sqrt{2Jm\omega \hbar }\left[ \sin \gamma +\tau \left[
\gamma \cos \gamma -\frac{\sin \gamma }{2}+J\cos \gamma \left( \gamma -\frac{%
\sin 2\gamma }{2}\right) \right] \right] ,  \notag
\end{eqnarray}%
then follow trivially from (\ref{a}) and (\ref{ad}). Expanding $x^{2}$ and $%
p^{2}$ in terms of $a^{\dagger }$ and $a$, a similar, albeit more lengthy,
computation yields%
\begin{eqnarray}
\left\langle J,\gamma ,\phi \right\vert x^{2}\left\vert J,\gamma ,\phi
\right\rangle  &=&\frac{\hbar }{2m\omega }\left[ 1+4J\cos ^{2}\gamma -\tau
J\left( 6\gamma \sin 2\gamma +\cos 2\gamma +2\right) \right.   \label{x2} \\
&&\left. \qquad \quad -\tau J^{2}(4\gamma \sin 2\gamma -\cos 4\gamma +1)
\right] ,  \notag \\
\left\langle J,\gamma ,\phi \right\vert p^{2}\left\vert J,\gamma ,\phi
\right\rangle  &=&\frac{\hbar m\omega }{2}\left[ 1+4J\sin ^{2}\gamma +\tau
J\left( 6\gamma \sin 2\gamma +\cos 2\gamma -2\right) \right.   \label{p2} \\
&&\left. \qquad \quad +\tau J^{2}(4\gamma \sin 2\gamma +\cos 4\gamma -1)
\right] .  \notag
\end{eqnarray}%
These two expressions may be used to compute the expectation value of $h$,
as defined in (\ref{h}), with regard to the GK-coherent states. The
remaining term in $h$ only needs to be computed to zeroth order to achieve
an overall accuracy of order $\tau $. We therefore calculate%
\begin{equation}
\left\langle J,\gamma ,\phi \right\vert
p^{2}x^{2}+x^{2}p^{2}+2xp^{2}x\left\vert J,\gamma ,\phi \right\rangle =\hbar
^{2}(1+4J+2J^{2}-2J^{2}\cos 4\gamma )+\mathcal{O}(\tau ).
\end{equation}%
Summing the contributions from (\ref{xp}), (\ref{x2}) and (\ref{p2}),
together with the required pre-factors to make up the Hamiltonian $h$,
yields the action identity (\ref{AI}) as expected. We remark that this
crucial identity was violated in \cite{GhoshRoy}.

Employing the above quantities we can also investigate how close the
coherent states approach the minimum uncertainty product. Assembling the
required expectation values we then obtain 
\begin{eqnarray}
\Delta x^{2} &=&\left\langle J,\gamma ,\phi \right\vert x^{2}\left\vert
J,\gamma ,\phi \right\rangle -\left\langle J,\gamma ,\phi \right\vert
x\left\vert J,\gamma ,\phi \right\rangle ^{2}=\frac{\hbar }{2m\omega }\left[
1+\tau J(\cos 2\gamma -2\gamma \sin 2\gamma )\right] ,~~~~~~ \\
\Delta p^{2} &=&\left\langle J,\gamma ,\phi \right\vert p^{2}\left\vert
J,\gamma ,\phi \right\rangle -\left\langle J,\gamma ,\phi \right\vert
p\left\vert J,\gamma ,\phi \right\rangle ^{2}=\frac{\hbar m\omega }{2}\left[
1-\tau J(\cos 2\gamma -2\gamma \sin 2\gamma )\right] ,
\end{eqnarray}%
and therefore 
\begin{equation}
\Delta x\Delta p=\frac{\hbar }{2}+\mathcal{O}(\tau ^{2}).
\end{equation}%
Thus the states $\left\vert J,\gamma ,\phi \right\rangle $ saturate the
minimal uncertainty in a simultaneous measurement of $x$ and $p$ and
therefore constitute squeezed states for all values of $J$ and $\gamma $ up
to first order in perturbation theory.

Using (\ref{xp}) we also verify Ehrenfest's theorem (\ref{Ehren}) for the
operators $x$ 
\begin{eqnarray}
&&i\hbar \frac{d}{dt}\left\langle J,\gamma +t\omega ,\phi \right\vert
x\left\vert J,\gamma +t\omega ,\phi \right\rangle =\left\langle J,\gamma
+t\omega ,\phi \right\vert [x,h]\left\vert J,\gamma +t\omega ,\phi
\right\rangle , \\
&=&\left\langle J,\gamma +t\omega ,\phi \right\vert \frac{i\hbar }{m}p+\frac{%
i\tau \omega }{2}(px^{2}+x^{2}p+2xpx)\left\vert J,\gamma +t\omega ,\phi
\right\rangle  \notag \\
&=&-i\hbar ^{3/2}\sqrt{\frac{2J\omega }{m}}\left[ \sin \hat{\gamma}+\tau %
\left[ \frac{1}{2}\sin \hat{\gamma}+\cos \hat{\gamma}\left( (J+1)\hat{\gamma}%
+\frac{3}{2}J\sin 2\hat{\gamma}\right) \right] \right]  \notag
\end{eqnarray}%
and $p$%
\begin{eqnarray}
&&i\hbar \frac{d}{dt}\left\langle J,\gamma +t\omega ,\phi \right\vert
p\left\vert J,\gamma +t\omega ,\phi \right\rangle =\left\langle J,\gamma
+t\omega ,\phi \right\vert [p,h]\left\vert J,\gamma +t\omega ,\phi
\right\rangle , \\
&=&\left\langle J,\gamma +t\omega ,\phi \right\vert -i\hbar m\omega
^{2}x-i\tau \omega (px^{2}+x^{2}p)\left\vert J,\gamma +t\omega ,\phi
\right\rangle  \notag \\
&=&-i\sqrt{2Jm}\hbar ^{3/2}\omega ^{3/2}\left[ \cos \hat{\gamma}+\frac{\tau 
}{4}\left[ (3J+2)\cos \hat{\gamma}-4(J+1)\hat{\gamma}\sin \hat{\gamma}%
-3J\cos 3\hat{\gamma}\right] \right] .  \notag
\end{eqnarray}%
For convenience we abbreviated here $\hat{\gamma}:=\gamma +t\omega $.

\subsection{Obervables in the non-Hermitian system}

As stated, the system we actually wish to investigate is described by the
non-Hermitian Hamiltonian (\ref{HD1}) with a non-trivial commutation
relation (\ref{one}) for its associated observables $X$ and $P$. In order to
test the inequality (\ref{HU}) we need to compute 
\begin{eqnarray}
\left\langle J,\gamma ,\Phi \right\vert X\left\vert J,\gamma ,\Phi
\right\rangle _{\eta } &=&\sqrt{\frac{2J\hbar }{m\omega }}\left[ \cos \gamma
+\frac{\tau }{2}\sin \gamma (J\sin 2\gamma -2\gamma (1+J))\right] ,
\label{X} \\
\left\langle J,\gamma ,\Phi \right\vert X^{2}\left\vert J,\gamma ,\Phi
\right\rangle _{\eta } &=&\frac{\hbar }{2m\omega }\left[ 1+4J\cos ^{2}\gamma
+\tau \lbrack 1+J(2-2\cos 2\gamma -6\gamma \sin 2\gamma )\right. ~~~
\label{XX} \\
&&+\left. 2J^{2}\sin 2\gamma (\sin 2\gamma -2\gamma )]\right] .  \notag
\end{eqnarray}%
We note here that the actual computation has been carried out by translating
first all quantities to a Hermitian setting and then following the same
reasoning as in the previous subsection. Combining (\ref{X}) and (\ref{XX})
then yields 
\begin{eqnarray}
\Delta X^{2} &=&\left\langle J,\gamma ,\Phi \right\vert X^{2}\left\vert
J,\gamma ,\Phi \right\rangle _{\eta }-\left\langle J,\gamma ,\Phi
\right\vert X\left\vert J,\gamma ,\Phi \right\rangle _{\eta }^{2}
\label{DXX} \\
&=&\frac{\hbar }{2m\omega }\left[ 1+\tau \left( 1+J(2-2\gamma \sin 2\gamma
-\cos 2\gamma )\right) \right] .  \notag
\end{eqnarray}%
The computation for the expectation values of $P$ is simpler, since the
metric commutes with $p$, such that%
\begin{equation}
\left\langle J,\gamma ,\Phi \right\vert P\left\vert J,\gamma ,\Phi
\right\rangle _{\eta }=\left\langle J,\gamma ,\phi \right\vert p\left\vert
J,\gamma ,\phi \right\rangle ,~~\text{and~~~}\left\langle J,\gamma ,\Phi
\right\vert P^{2}\left\vert J,\gamma ,\Phi \right\rangle _{\eta
}=\left\langle J,\gamma ,\phi \right\vert p^{2}\left\vert J,\gamma ,\phi
\right\rangle ,
\end{equation}%
and therefore%
\begin{equation}
~~\Delta P^{2}=\Delta p^{2}.  \label{DP}
\end{equation}%
Expanding finally (\ref{DXX}) and (\ref{DP}), we obtain%
\begin{equation}
\Delta X\Delta P=\frac{\hbar }{2}\left[ 1+\frac{\tau }{2}\left( 1+4J\sin
^{2}\gamma \right) \right] =\frac{\hbar }{2}\left( 1+\hat{\tau}\left\langle
J,\gamma ,\Phi \right\vert P^{2}\left\vert J,\gamma ,\Phi \right\rangle
\right) .
\end{equation}%
This means that also in the non-Hermitian setting the minimal uncertainty
product for the observables $X$ and $P$, commuting as specified in (\ref{one}%
), is saturated. Thus to first order in perturbation theory also the
GK-coherent states $\left\vert J,\gamma ,\Phi \right\rangle $ are squeezed
states, remarkably this holds irrespective of the values for $J$ and $\gamma 
$.

Apparently this result was also obtained in \cite{GhoshRoy}, but our
disagreement with the results presented in there is at least fourfold.
Firstly, the authors used the incorrect representation for the canonical
variables $X$ and $P$ as mentioned earlier. Secondly the authors computed
conceptually the wrong expectations values even when using their
representation. Thirdly, the authors only take the first order in (\ref{phi}%
) into account and therefore miss out various terms contributing to the
first order calculation in $\tau $. Finally, even ignoring the previous
three points and adopting all the wrong concepts used in \cite{GhoshRoy}, we
disagree on a purely computational level with many of the expressions
presented in there.

Next we also verify Ehrenfest's theorem (\ref{Ehren}) for the operators $X$ 
\begin{eqnarray}
&&i\hbar \frac{d}{dt}\left\langle J,\gamma +t\omega ,\Phi \right\vert
X\left\vert J,\gamma +t\omega ,\Phi \right\rangle _{\eta }=\left\langle
J,\gamma +t\omega ,\Phi \right\vert [X,H]\left\vert J,\gamma +t\omega ,\Phi
\right\rangle _{\eta },  \label{E1} \\
&=&\left\langle J,\gamma +t\omega ,\Phi \right\vert \frac{i\hbar }{m}(P+%
\check{\tau}P^{3})\left\vert J,\gamma +t\omega ,\Phi \right\rangle _{\eta } 
\notag \\
&=&-i\hbar ^{3/2}\sqrt{\frac{2J\omega }{m}}\left[ \sin \hat{\gamma}+\tau %
\left[ (J+1)\hat{\gamma}\cos \hat{\gamma}+\frac{1}{2}\sin \hat{\gamma}%
(2+J-3J\cos 2\hat{\gamma})\right] \right]  \notag
\end{eqnarray}%
and the operator $P$%
\begin{eqnarray}
&&i\hbar \frac{d}{dt}\left\langle J,\gamma +t\omega ,\Phi \right\vert
P\left\vert J,\gamma +t\omega ,\Phi \right\rangle _{\eta }=\left\langle
J,\gamma +t\omega ,\Phi \right\vert [P,H]\left\vert J,\gamma +t\omega ,\Phi
\right\rangle _{\eta },  \label{E2} \\
&=&\left\langle J,\gamma +t\omega ,\Phi \right\vert -im\hbar \omega
^{2}\left( X+\frac{\check{\tau}}{2}XP^{2}+\frac{\check{\tau}}{2}%
P^{2}X\right) \left\vert J,\gamma +t\omega ,\Phi \right\rangle _{\eta } 
\notag \\
&=&-i\sqrt{2Jm}\hbar ^{3/2}\omega ^{3/2}\left[ \cos \hat{\gamma}+\frac{\tau 
}{4}\left[ (3J+2)\cos \hat{\gamma}-4(J+1)\hat{\gamma}\sin \hat{\gamma}%
-3J\cos 3\hat{\gamma}\right] \right] .  \notag
\end{eqnarray}%
Taking now for simplicity $\gamma =0$, differentiating (\ref{E1}) once again
and combining it with (\ref{E2}) we obtain the corresponding identity to
Newton's equation of motion%
\begin{equation}
\left\langle J,t\omega ,\Phi \right\vert \ddot{X}\left\vert J,t\omega ,\Phi
\right\rangle _{\eta }=-\omega ^{2}\left\langle J,t\omega ,\Phi \right\vert
X+\frac{\check{\tau}}{2}(3XP^{2}+3P^{2}X+2PXP)\left\vert J,t\omega ,\Phi
\right\rangle _{\eta },  \label{Newton}
\end{equation}%
The relations (\ref{E1}) and (\ref{E2}) were not recovered in \cite{GhoshRoy}%
, where the comparison between the left and right hand sides mismatched.
Instead of (\ref{Newton}) the authors proposed a "correspondence principle
with twist". According to our argumentation this is incorrect and there is
in fact no reason to assume the Newton's equation is simply the same as the
one for the standard harmonic oscillator. The reason for the discrepancy are
the aforementioned conceptual and computational mistakes in \cite{GhoshRoy}.

\section{Fractional revival structure}

Further insights into the comparison between the classical and quantum
description can be obtained from the revival time for wave packets. For a
general wave packet of the form $\psi =\dsum c_{n}\phi _{n}$ sufficiently
well localized, i.e. being governed by sub-Poissonian statistics, near a
submode $n=\bar{n}$ with energy $E_{\bar{n}}$, it was argued in \cite{Perel}
that beside the revival of the classical-like wave packet after the
classical period $T_{cl}=2\pi \hbar /\left\vert E_{\bar{n}}^{\prime
}\right\vert $ one may also encounter so-called fractional revivals. These
partial revivals of the original wave packet occur at times $p/q(T_{rev})$,
with coprime integers $p$, $q$ and the revival time given by $T_{rev}=4\pi
\hbar /\left\vert E_{\bar{n}}^{\prime \prime }\right\vert $. Depending on
the values of $p$ and $q$ one can interpret the emerging features as
different types of superpositions of classical-like wave packets.

For the case at hand we may follow \cite{Klauder2} and expand the energy $%
E_{n}$ in the expression for the wave packets (\ref{GK}) 
\begin{equation}
\left\vert J,\omega t,\phi \right\rangle =\sum\limits_{n=0}^{\infty
}c_{n}(J)\exp (-itE_{n}/\hbar )\left\vert \phi _{n}\right\rangle ,
\label{sum}
\end{equation}%
with weighting function $c_{n}(J)=J^{n/2}/\mathcal{N}(J)\sqrt{\rho _{n}}$,
about $\bar{n}:=\left\langle n\right\rangle =Jd\ln \mathcal{N}^{2}(J)/dJ$.
To first order in perturbation theory we then easily compute%
\begin{equation}
\left\langle n\right\rangle =J-\tau \left( J+\frac{J^{2}}{2}\right) +%
\mathcal{O}(\tau ^{2}),\quad \text{and\quad }\left\langle n^{2}\right\rangle
=J+J^{2}-\tau \left( J+3J^{2}+J^{3}\right) +\mathcal{O}(\tau ^{2}),
\label{nb}
\end{equation}%
such that%
\begin{equation}
\Delta n^{2}=\left\langle n^{2}\right\rangle -\left\langle n\right\rangle
^{2}=J-\tau \left( J+J^{2}\right) +\mathcal{O}(\tau ^{2}).
\end{equation}%
Consequently the Mandel parameter $Q$ \cite{Mandel} turns out to be negative%
\begin{equation}
Q:=\frac{\Delta n^{2}}{\left\langle n\right\rangle }-1=-\frac{J\tau }{2}+%
\mathcal{O}(\tau ^{2})<0,
\end{equation}%
suggesting a sub-Poissonian statistics. This implies that we have a strong
localization around $\bar{n}$ required for the possibility of the revival of
the classical-like sub-wave packet. According to the above expressions we
compute the classical period and the revival time to 
\begin{equation}
T_{\text{cl}}=\frac{2\pi }{\omega }-\frac{\tau }{\omega }(1+2J)\pi ,\quad
\quad \text{and\quad \quad }T_{\text{rev}}=\frac{4\pi }{\omega \tau }.
\label{TT}
\end{equation}%
We now use these quantities to analyze the behaviour of the autocorrelation
function%
\begin{equation}
A(t):=\left\vert \left\langle J,\gamma ,\phi \right. \left\vert J,\gamma
+t\omega ,\phi \right\rangle \right\vert ^{2}=\left\vert \left\langle
J,\gamma ,\Phi \right. \left\vert J,\gamma +t\omega ,\Phi \right\rangle
_{\eta }\right\vert ^{2}.
\end{equation}%
In order to find a set of meaningful values for our free parameters $J$, $%
\tau $ and also to find an appropriate upper limit cutoff in the sum (\ref%
{sum}), let us first investigate the weighting function $c_{n}(J)$.

\begin{figure}[h]
\centering   \includegraphics[width=7.5cm,height=6.0cm]{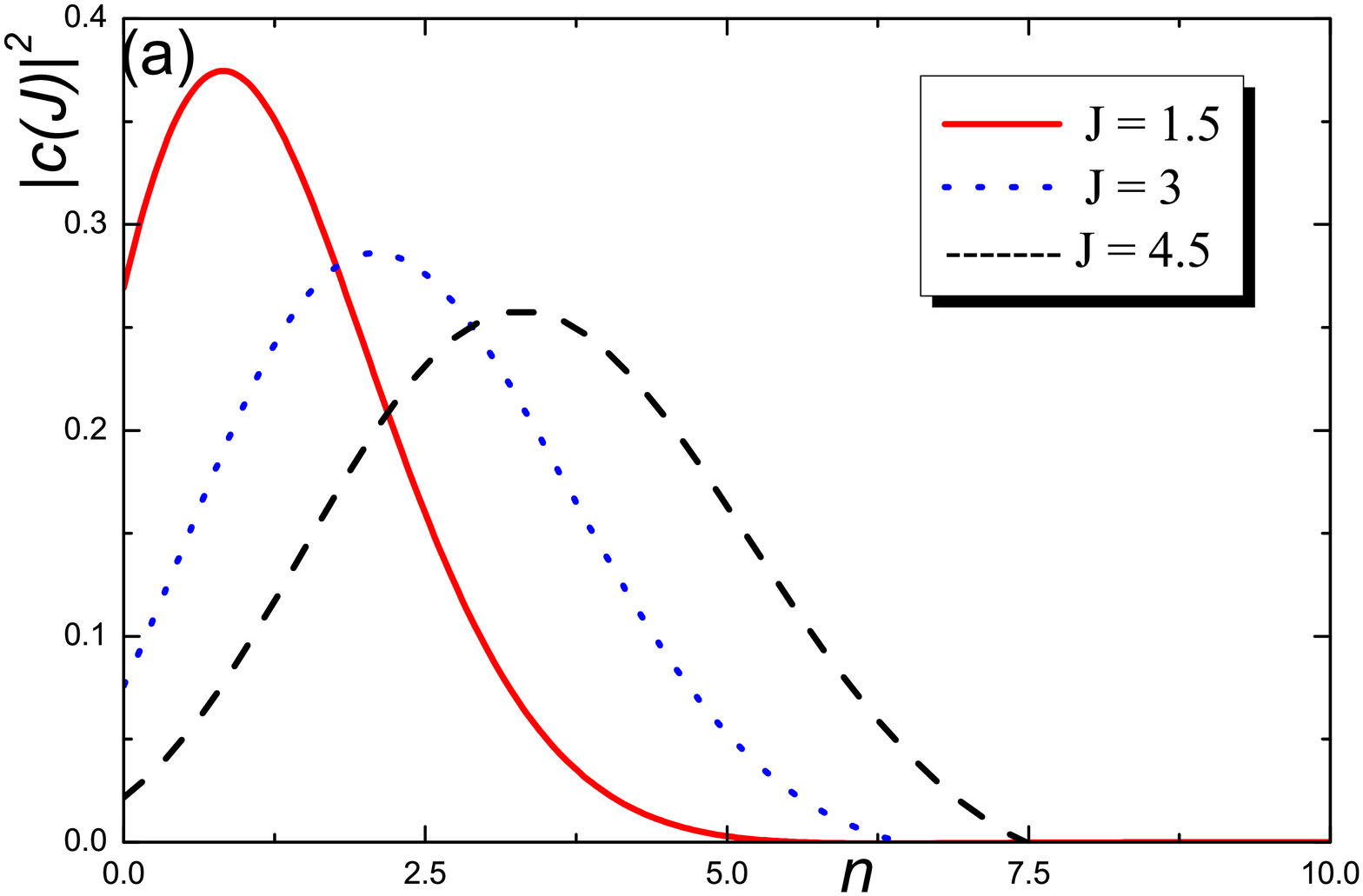} %
\includegraphics[width=7.5cm,height=6.0cm]{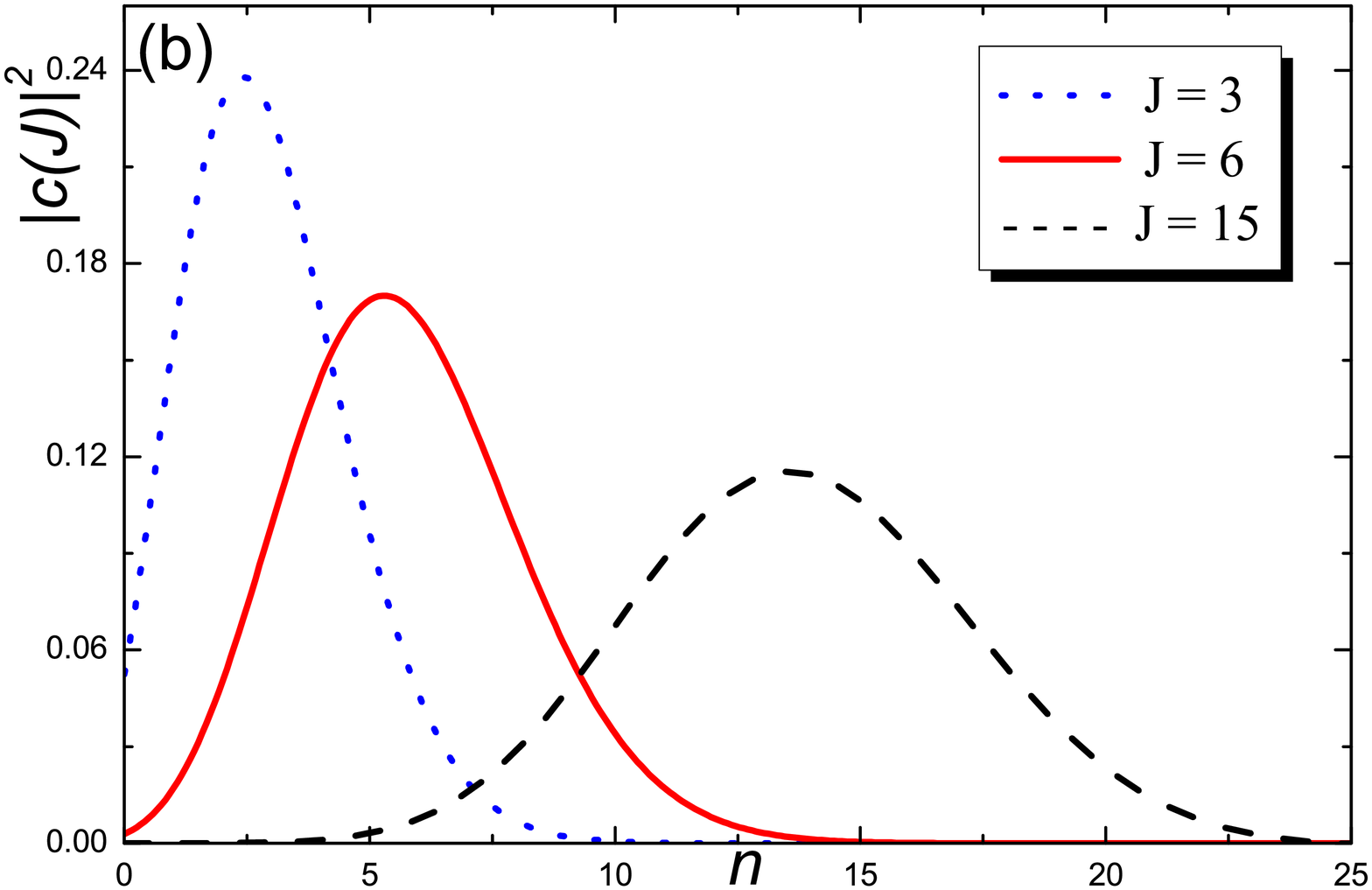}
\caption{(a) Weighting function for $\protect\tau =0.1$ with $\left\langle
n\right\rangle =1.24,2.25,3.04$ for $J=1.5,3,4.5$, respectively and (b) $%
\protect\tau =0.01$ with $\left\langle n\right\rangle =2.93,5.76,13.72$ for $%
J=3,6,15$, respectively.}
\label{F2}
\end{figure}
\begin{figure}[h!]
\centering   \includegraphics[width=7.5cm,height=6.0cm]{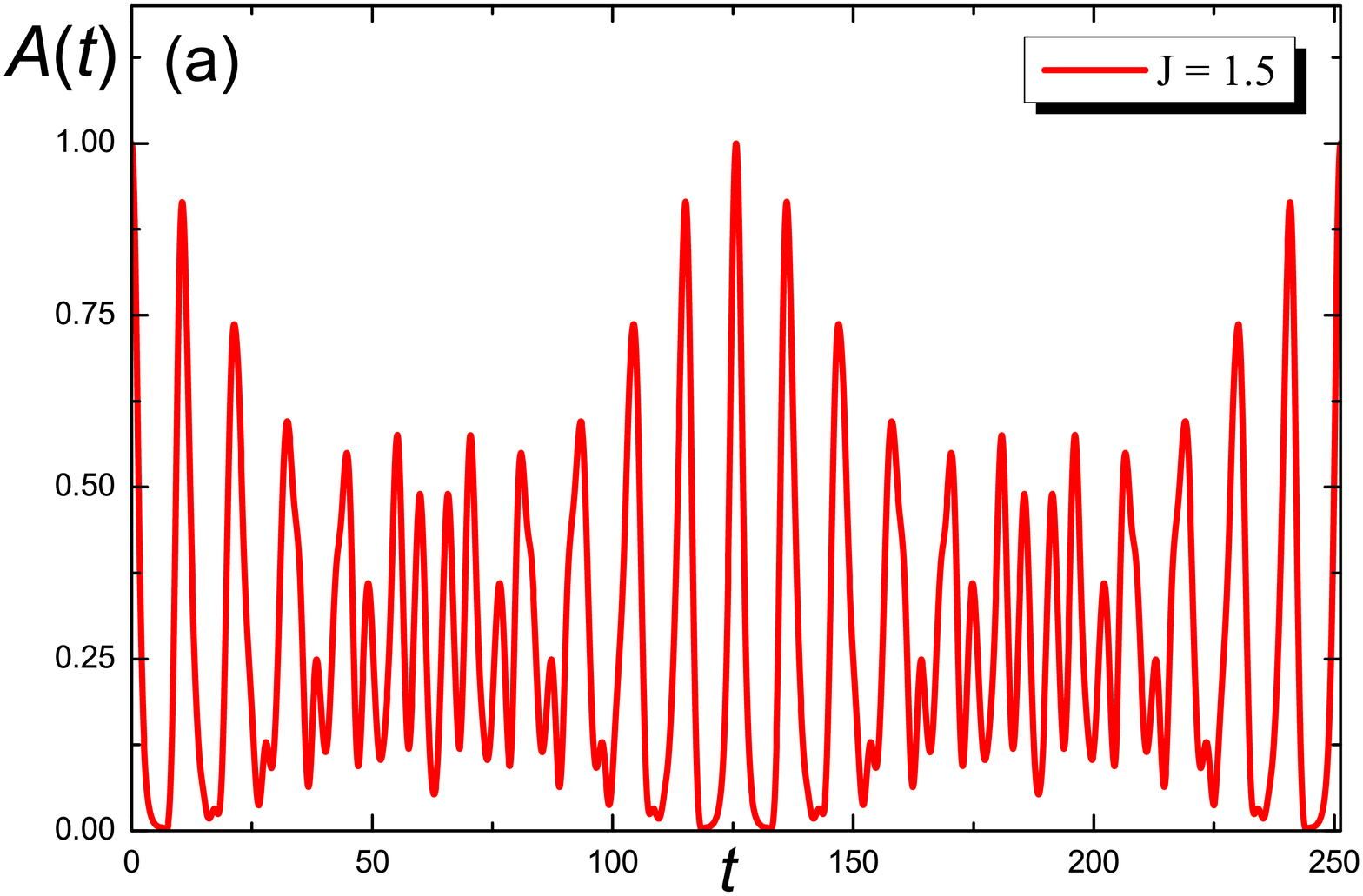} %
\includegraphics[width=7.5cm,height=6.0cm]{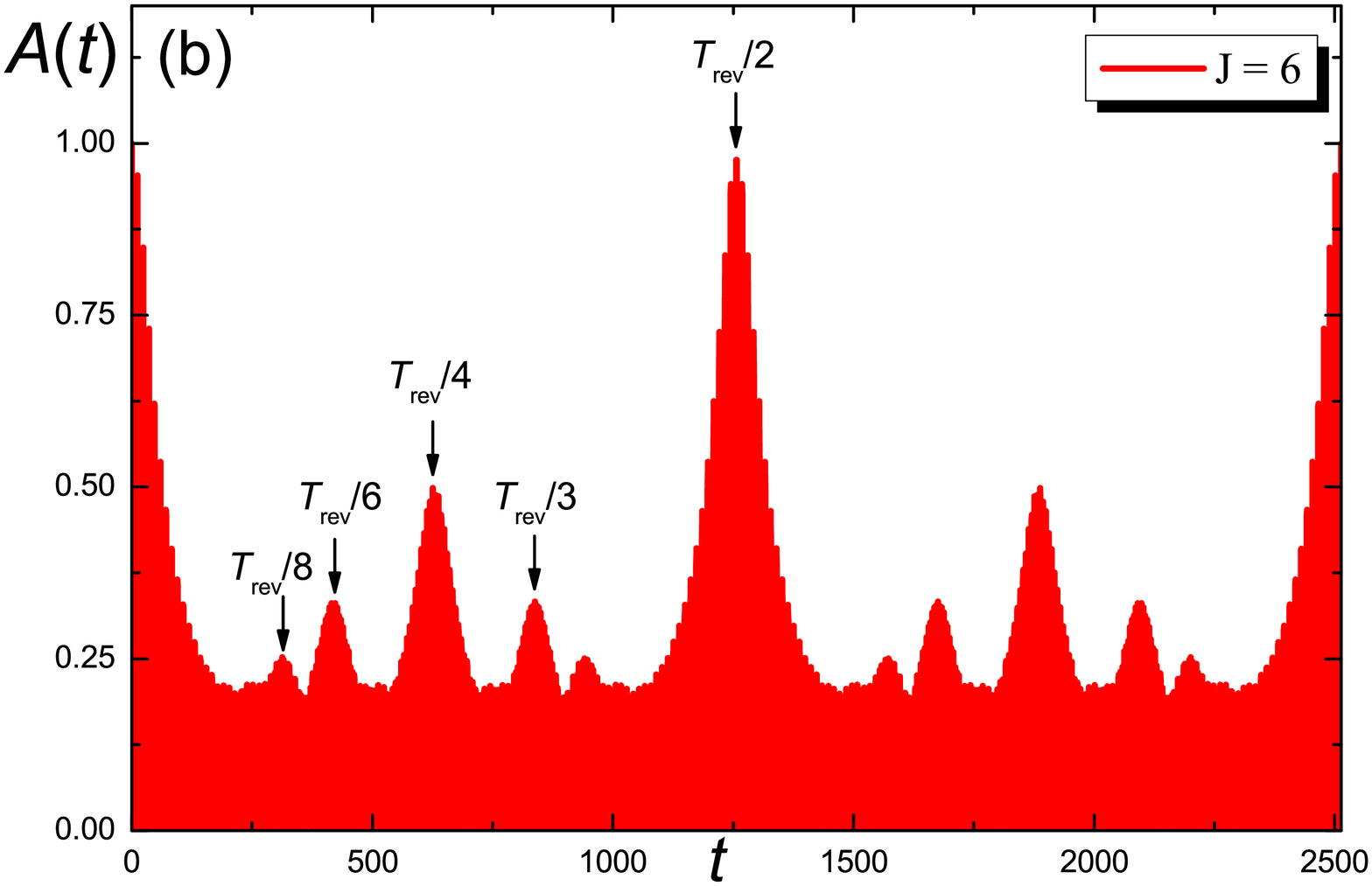}
\caption{(a) Autocorrelation function as a function of time for $J=1.5$, $%
\protect\tau =0.1$, $\protect\omega =0.5$, $\hbar =1$, $\protect\gamma =0$, $%
T_{\text{cl}}=10.05$ and $T_{\text{rev}}=251.32$; (b) Autocorrelation
function as a function of time for $J=6$, $\protect\tau =0.01$, $\protect%
\omega =0.5$, $\hbar =1$, $\protect\gamma =0$, $T_{\text{cl}}=11.74$ and $T_{%
\text{rev}}=2513.27$.}
\label{F1}
\end{figure}

For the chosen values we observer in figure \ref{F2} that the wave packets
are well localized around $\bar{n}$ resulting from (\ref{nb}), such that the
prerequisite for the validity of the analysis in \cite{Perel} is given.
Increasing the values of $J$ for fixed $\tau $ we observe negative values
for $|c(J)|^{2}$ for large values of $n$, which clearly indicates that our
pertubative expressions are no longer valid in that regime. We also note
that $n\approx 50$ will be a sufficiently good value to terminate the sum in
the expression for the autocorrelation function (\ref{sum}) analyzed in
figure \ref{F1}.

In panel (a) of figure \ref{F1} we clearly observe local maxima at multiples
of the classical period $T_{\text{cl}}$. As explained in \cite{Perel} the
first full reconstruction of the original wave packet is obtained at $T_{%
\text{rev}}/2$ which is clearly visible in panel (a). The fractional
revivals are better observed for smaller values of $\tau $ as depicted in
panel (b). In that scenario the classical periods are so small as compared
to the revival time that they are no longer resolved. We clearly observe a
number of fractional revivals, such as for instance $T_{\text{rev}}/4$
corresponding to the superposition of two classical-like sub-wave packets
and others as indicated in the figure.

Notice that our expression and our analysis presented here differs once
again from the one in \cite{GhoshRoy}, where for instance the mandatory
revival at $T_{\text{rev}}/2$ was not observed.

\section{Conclusions}

Our central results is the construction of explicit expressions for the
GK-coherent states for a non-Hermitian system on a noncommutative space
leading to a generalized version of Heisenberg's uncertainty relation. We
showed that these states are squeezed for all values of $J$ and $\gamma $ as
they saturate the minimal uncertainty. Crucially we established two of the
nontrivial GK-axioms are satisfied. First of all the states are shown to be
temporarily stable, i.e. they remain coherent under time evolution, and
second the states satisfy the action identity (\ref{AI}) allowing for a
close relation to a classical description in terms of action angle
variables. We also demonstrated that when using the appropriate metric
Ehrenfest's theorem is satisfied for the observables $X$ and $P$. The
desired resemblance of the coherent states with a classical description was
further underpinned by an analysis of the revival structure exhibiting the
typical quasiclassical evolution of the original wave packet. It should be
noted that in the considered case the wave packet revival time (\ref{TT})
depends explicitly on the deformation parameter $\tau $, such that a
possible measurement could distinguish between a noncommutative and a
standard commutative space. For instance, in the order of femtoseconds half
and quarter revivals have been observed experimentally \cite{Mol} for
molecular wave packets described by anharmonic oscillator potentials with
eigenenergies similar to (\ref{En}). Our analysis holds to first order
perturbation theory in $\tau $ and of course it would be very interesting to
extend this to higher order or eventually to the exact case.

There are various other directions into which our analysis might be taken
forward. For instance, different types of models might exhibit a varied
behaviour. More challenging is a construction for such states in higher
dimensions. A systematic comparison with different types of coherent states
would be insightful, especially with rare constructions related to
non-Hermitian Hamiltonians \cite{Evacom}.\medskip

\noindent \textbf{Acknowledgments:} SD is supported by a City University
Research Fellowship. AF thanks Carla Faria for useful comments and
discussions.

\appendix

\section{Appendix}

We present here a sample computation in order to make our results
transparent and reproducible. In addition, we highlight the differences
compared to \cite{GhoshRoy}. One of the simplest computation in this context
is to evaluate the expectation value of the annihilation operator (\ref{a}).
The evaluation of expectation values for different types of operators is
more involved, but goes along the same lines. Using the expression for (\ref%
{GK}) we compute%
\begin{equation}
\left\langle J,\gamma ,\phi \right\vert a\left\vert J,\gamma ,\phi
\right\rangle =\frac{1}{\mathcal{N}^{2}}\sum\limits_{n,m=0}^{\infty }\frac{%
J^{(m+n)/2}\exp \left[ i\gamma (e_{m}-e_{n})\right] }{\sqrt{\rho _{m}\rho
_{n}}}\left\langle \phi _{m}\right\vert a\left\vert \phi _{n}\right\rangle .
\end{equation}%
With the expansion of $\left\vert \phi _{n}\right\rangle $ to first order in 
$\tau $ we obtain 
\begin{eqnarray}
\left\langle \phi _{m}\right\vert a\left\vert \phi _{n}\right\rangle &=&%
\sqrt{n}\delta _{m,n-1}+\frac{\tau }{16}\left( \sqrt{(n+1)_{4}}\sqrt{n}-%
\sqrt{(n-3)_{4}}\sqrt{n-4}\right) \delta _{m,n-5} \\
&&~~~~~~~~~~~~~\ +\frac{\tau }{16}\left( \sqrt{(n+1)_{4}}\sqrt{n+4}-\sqrt{%
(n-3)_{4}}\sqrt{n}\right) \delta _{m,n+3},  \notag \\
&=&\sqrt{n}\delta _{m,n-1}+\frac{\tau }{4}\sqrt{(n+1)(n+2)(n+3)}\delta
_{m,n+3},  \notag
\end{eqnarray}%
such that%
\begin{equation}
\left\langle J,\gamma ,\phi \right\vert a\left\vert J,\gamma ,\phi
\right\rangle =\sum\limits_{n=1}^{\infty }\frac{\sqrt{n}J^{n-1/2}e^{i\gamma
(e_{n-1}-e_{n})}}{\mathcal{N}^{2}\sqrt{\rho _{n-1}\rho _{n}}}+\tau
\sum\limits_{n=0}^{\infty }\frac{J^{n+3/2}\sqrt{(n+1)_{3}}e^{i\gamma
(e_{n+3}-e_{n})}}{4\mathcal{N}^{2}\sqrt{\rho _{n+3}\rho _{n}}}+\mathcal{O}%
(\tau ^{2}).  \label{a3}
\end{equation}%
The last sum has been ignored in \cite{GhoshRoy}, but is an important
contribution to order $\tau $. Using $e_{n-1}-e_{n}=-1-n\tau $ and $\rho
_{n}=\rho _{n-1}e_{n}$ the first sum in (\ref{a3}) is evaluated as%
\begin{eqnarray}
&&\frac{e^{-i\gamma }}{\mathcal{N}^{2}}\sum\limits_{n=1}^{\infty }\frac{%
J^{n-1/2}e^{-i\gamma n\tau }}{\rho _{n-1}\sqrt{1+\frac{\tau }{2}(1+n)}}=%
\frac{e^{-i\gamma }}{\mathcal{N}^{2}}\sum\limits_{n=1}^{\infty }\frac{%
J^{n-1/2}}{\rho _{n-1}}\left[ 1-\frac{\tau }{4}(1+n+4i\gamma n)\right] +%
\mathcal{O}(\tau ^{2}),~~~~~  \label{a5} \\
&=&\frac{e^{-i\gamma }}{\sqrt{J}}\left[ \left( 1-\frac{\tau }{4}\right) 
\frac{\sum\limits_{n=1}^{\infty }\frac{J^{n}}{\rho _{n-1}}}{%
\sum\limits_{n=0}^{\infty }\frac{J^{n}}{\rho _{n}}}-\frac{\tau }{4}%
(1+4i\gamma )\frac{\sum\limits_{n=1}^{\infty }\frac{nJ^{n}}{\rho _{n-1}}}{%
\sum\limits_{n=0}^{\infty }\frac{J^{n}}{\rho _{n}}}\right] +\mathcal{O}(\tau
^{2}), \\
&=&\sqrt{J}e^{-i\gamma }\left[ 1-\frac{\tau }{4}\left( 2+J+4i\gamma
(1+J)\right) \right] +\mathcal{O}(\tau ^{2}).  \label{a6}
\end{eqnarray}%
For the second sum in (\ref{a3}) we use $e_{n+3}-e_{n}=3+3\tau (2+n)$ and $%
\rho _{n+3}=\rho _{n}e_{n+3}e_{n+2}e_{n+1}$, such that it becomes%
\begin{equation}
\tau \sum\limits_{n=0}^{\infty }\frac{J^{n+3/2\sqrt{(n+1)_{3}}}e^{i\gamma
(3+3\tau (2+n))}}{4\mathcal{N}^{2}\rho _{n}\sqrt{e_{n+3}e_{n+2}e_{n+1}}}+%
\mathcal{O}(\tau ^{2})=\frac{\tau J^{3/2e^{3i\gamma }}}{4\mathcal{N}^{2}}%
\sum\limits_{n=0}^{\infty }\frac{J^{n}}{\rho _{n}}+\mathcal{O}(\tau ^{2})=%
\frac{\tau J^{3/2e^{3i\gamma }}}{4}+\mathcal{O}(\tau ^{2}).  \label{a7}
\end{equation}%
Collecting (\ref{a6}) and (\ref{a7}) we obtain (\ref{a}). Similarly we
compute the expectation values for $x^{2}$, $p^{2}$, $x^{2}p^{2}$ etc by
converting them first into expressions involving the $a$ and $a^{\dagger }$
and then using the arguments from above. For the noncommutative scenario we
convert first to a setting involving Hermitian operators. For instance, we
use 
\begin{equation}
\left\langle J,\gamma ,\Phi \right\vert X\left\vert J,\gamma ,\Phi
\right\rangle _{\eta }=\left\langle J,\gamma ,\phi \right\vert x+\frac{%
\check{\tau}}{2}(p^{2}x+xp^{2})\left\vert J,\gamma ,\phi \right\rangle +%
\mathcal{O}(\tau ^{2})
\end{equation}%
and compute the right hand side as explained above.

\newif\ifabfull\abfulltrue


\end{document}